\documentclass[aps,prl,showpacs,nofootinbib,twocolumn,float,groupedaddress]{revtex4}
\usepackage{amsmath,bm,epsfig,graphicx}

\begin{document}
\title{A novel route to Bose-Einstein condensation of two-electron atoms}
\author{Purbasha Halder, Chih-Yun Yang, and Andreas Hemmerich}
\affiliation{Institut f\"{u}r Laserphysik, Universit\"{a}t Hamburg, Luruper Chaussee 149, 22761 Hamburg, Germany}

\begin{abstract}
We present a novel route to Bose-Einstein condensation devised for two-electron atoms, which do not admit practicable cooling techniques based upon narrow intercombination lines. A dipole trap for $^{40}$Ca atoms in the singlet ground state is loaded from a moderately cold source of metastable triplet atoms via spatially and energetically selective optical pumping permitting four orders of magnitude increase of the phase space density. Further cooling to quantum degeneracy is achieved by forced evaporation optimized to minimize three-body losses. In a combined loading and evaporation cycle of less than three seconds we are able to condense 3000 atoms.
\end{abstract}

\date{\today}

\maketitle
Owing to their unique electronic properties, alkaline-earth-metal (AEM) atoms and certain rare-earth (RE) atoms are at the focus of several rapidly evolving lines of research in atomic physics. The existence of narrow intercombination lines connecting the non-magnetic singlet state to the triplet states render them highly useful as atomic frequency standards \cite{Did2004, Je2008, Lem2009, Mar2009}. These narrow transitions can also serve as novel precision tools for read-out and manipulation of quantum information in optical lattices \cite{Gor2009} or for the control of interaction strength in the quantum degenerate regime \cite{Ciu2005, Nai2006}. Ultracold fermionic AEM samples trapped in optical lattices provide intriguing new possibilities for simulating electronic matter dominated by spin-orbit coupling \cite{Gor2010}. Bosonic AEM isotopes with singlet and long-lived metastable triplet states uncomplicated by hyperfine structure lend themselves for ultracold collision studies confronting theory at a fundamental level \cite{Wei1999, Zin2000, Bur2002, Der2003, Kok03, Han06}.

With such prospects ahead, laser cooling and trapping techniques for two-electron atoms have developed rapidly during the last decade. Unlike alkali metals, bosonic AEM and RE atoms have non-magnetic singlet ground states. This precludes the application of two key techniques commonly used in quantum gas preparation: sub-Doppler cooling techniques in ground state magneto-optical traps and evaporative cooling in magnetic traps. Reaching ultracold temperatures for AEM and RE atoms therefore relies on Doppler-cooling followed by evaporation in optical traps. Efficient Doppler-cooling, however, requires intercombination transitions whose spectral width is sufficiently broad to permit significant radiation pressure while being narrow enough to allow for low temperatures (see, for example, \cite{Kat1999}). To further complicate matters, for such narrow-line cooling to be effective in combination with optical dipole traps, the use of so called magic wavelengths - where equal optical potentials arise for the ground state and the excited state \cite{magic wavelength} - are required. Strontium and ytterbium have turned out to be ideal systems in this respect, which is documented by the fact that all stable isotopes of strontium \cite{Ste2009, Esc2009, Mic2010, Ste2010} and four of the five stable bosonic isotopes of ytterbium \cite{Yb2003, Yb2007, Yb2009,Yb2011} have been cooled down to form a Bose-Einstein condensate (BEC). In contrast to strontium and ytterbium, the intercombination lines of the two lightest AEM species - magnesium and calcium - are too narrow to be used for Doppler cooling, unless sophisticated line quenching and laser bandwidth shaping techniques are employed \cite{Bin2001, Deg2005}. Hence, approaching quantum degeneracy for these species is particularly challenging. Only a single research group has claimed the observation of a calcium BEC to date \cite{Kra2009}, while cooling of magnesium beyond the millikelvin range does not seem to be within reach yet, despite its possible relevance for metrology \cite{Fri2008, Jen2011}. An additional complication on the road towards a BEC of calcium is its large $s$-wave scattering length exceeding $400\,a_{0}$ ($a_{0}=$ Bohr radius). This gives rise to large three-body recombination losses and requires careful density management in evaporative cooling protocols.

\begin{figure}
\centering
\includegraphics[scale=0.35, angle=0, origin=c]{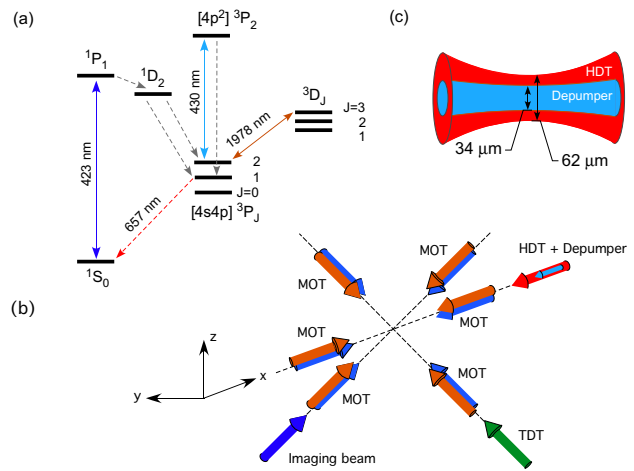}
\caption{(Color online) (a) Electronic level scheme of the calcium atom showing the transitions relevant to this work. Solid lines denote laser transitions, while dashed lines stand for spontaneous decay. (b) Orientation of the singlet- and triplet-MOT beams (blue and orange arrows), HDT beam (red arrow) and TDT beam (green arrow). Gravity acts along the $z$-axis. (c) Expanded view of the overlap between the horizontal dipole trap and depumper beams.}
\vspace{-6.mm}
\label{lasers_beams}
\end{figure}

\begin{figure}
\centering
\includegraphics[scale=0.25, angle=0, origin=c]{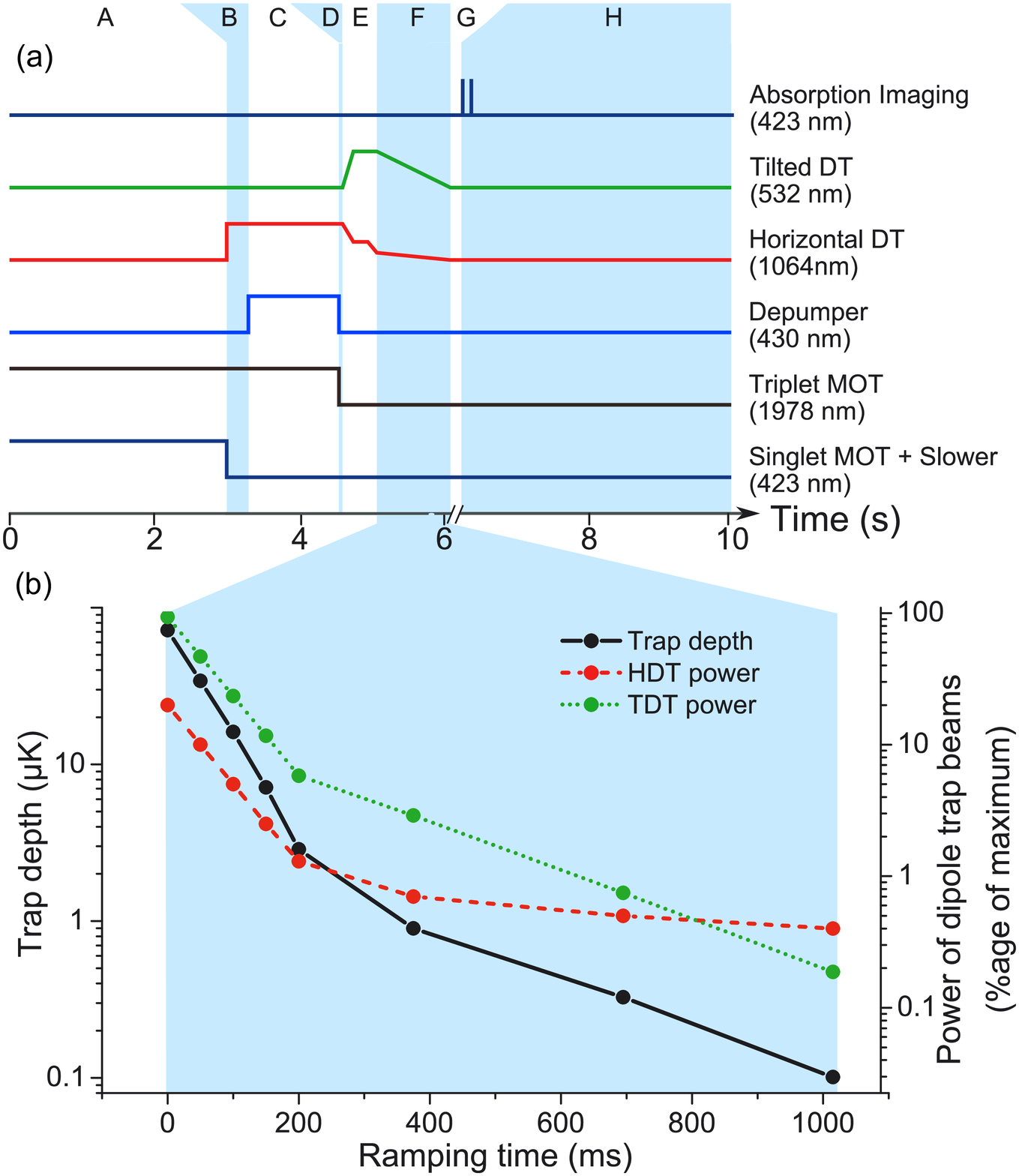}
\caption{ (Color online) (a) A typical experimental run comprises the following stages: A. loading of the triplet-MOT, B. cooling of the triplet-MOT, C. loading of the HDT, D. spontaneous evaporation in the HDT, E. loading of the CDT, F. Forced evaporation in the CDT, G. ballistic expansion of the atomic cloud for variable times after their release from the trap, H. imaging of the atomic cloud and processing of the image. The shown lines illustrate the powers of the laser beams indicated on the right. (b) Evaporation ramp applied in stage F: the solid black line indicates the trap depths in $\mu$K (left $y$-axis). The dashed red (dotted green) line shows the power in the HDT (TDT) beam (right $y$-axis).}
\vspace{-6.mm}
\label{Timing}
\end{figure}

In this letter we demonstrate a novel route to Bose-Einstein condensation of AEM or RE atoms, which does not require complex narrow-line cooling techniques, the use of magic wavelength optical traps or extensive laser bandwidth shaping. The inherent robustness and unspecific character of our scheme should permit its application in a wide range of experimental circumstances. In principle, our method can be used to prepare high-density ultracold AEM samples in any state with sufficiently long lifetime, provided these states can be loaded via optical pumping. This includes, for example, the singlet ground states of calcium and magnesium, but also long-lived excited states could be addressed. Here, we apply the scheme to prepare a BEC of calcium ($^{40}$Ca) in the singlet ground state.

Our experimental protocol proceeds in three steps: 1.\,conventional preparation of a laser cooled atomic sample in the triplet system; 2.\,continuous loading of a dipole trap for singlet ground state atoms via spatially selective optical pumping; 3.\,forced evaporation in the ground state dipole trap. As can be seen in Fig.~\ref{lasers_beams}(a), the principal fluorescence line of the singlet system, $^{1}$S$_{0} \rightarrow ^{1}$P$_{1}$ at 423\,nm, has a weak decay channel to the metastable $^{3}$P$_{2}$ state. In step \,1, we therefore use this transition to operate a Zeeman slower and a magneto-optical trap (MOT) (referred to as singlet-MOT) in order to load a second MOT (referred to as triplet-MOT) using the infrared closed cycle transition $^{3}$P$_{2} \rightarrow ^{3}$D$_{3}$ at 1978\,nm. The details are described in \cite{Gru2002, Han2003}. The loading is terminated by switching off the 423\,nm light for the Zeeman slower and the singlet-MOT beams. Further cooling of the triplet-MOT is then carried out by increasing the negative detuning of the infrared light by another 5\,MHz, yielding a sample of 2.3$\times$10$^{8}$ atoms at a temperature of 250\,$\mu$K (Fig.~\ref{Timing}(a), B). 

In step\,2 the atoms are loaded into a dipole trap oriented perpendicular to the direction of gravity (the $x$-axis in Fig.~\ref{lasers_beams}(b)), which is referred to as the horizontal dipole trap (HDT). The HDT beam is obtained from a 20\,W fibre laser (IPG photonics, YLR series) at 1064\,nm. With a beam waist of 31\,$\mu$m and 13.5\,W of power inside the vacuum chamber, this results in a maximum trap depth of 370\,$\mu$K for ground state atoms. This trap is loaded by optically pumping those $^{3}\textrm{P}_{2}$ atoms in the triplet MOT, which reside close to the potential minimum of the HDT. The spatially selective optical pumping is obtained from a weak (few hundred picowatts) laser beam (referred to as depumping beam) resonant with the $\,[4s4p]^{3}$P$_{2} \rightarrow \,[4p^2]^{3}$P$_{2}$ transition at 430\,nm. Its beam waist is 17\,$\mu$m and it is aligned to propagate well within the HDT beam (see Fig.~\ref{lasers_beams}(b) and (c)). The transfer of the atoms proceeds via spontaneous decay to $\,[4s4p]^{3}$P$_{1}$ and subsequently to $^{1}$S$_{0}$ as illustrated in Fig.~\ref{lasers_beams}(a). The low power and small spatial width of the depumping beam ensures that only the slowest atoms stay long enough within the illuminated volume to be pumped to the ground state and, hence, to be captured at the center of the HDT. As long as the depumping beam is active, accumulation of atoms in the dipole trap continues until loading is compensated by losses due to evaporation. The loading stage (Fig.~\ref{Timing}(a), C) continues for 1.25\,s and is terminated by switching off the optical pumping and triplet-MOT beams. After 100\,ms of spontaneous evaporation in the HDT we obtain 1.5$\times$10$^{6}$ atoms at 26\,$\mu$K with a density of 1.2$\times$10$^{12}$\,cm$^{-3}$. The corresponding phase space density (PSD) of 2$\times$10$^{-4}$ represents an almost four orders of magnitude increase over the PSD in the triplet-MOT. A quantitative theoretical study of the loading scheme can be found in our previous work \cite{Yan2007}. For the experimental run to produce a BEC, a 50\,ms spontaneous evaporation stage is used (Fig.~\ref{Timing}(a), D).

In order to increase the atomic confinement, in step\,3, we add a second trapping beam tilted with respect to the direction of gravity by 45$^\circ$ and aligned in order to intersect the HDT beam perpendicularly (see Fig.~\ref{lasers_beams}(b)). This beam is referred to as the tilted dipole trap (TDT) beam, while the combined dipole trap arising from the presence of both beams is referred to as crossed-beam dipole trap (CDT). The TDT beam is obtained from a solid state frequency doubled laser system (Coherent Verdi V5) which delivers 5\,W at 532\,nm. The maximum power available inside the vacuum chamber is 2.7\,W and the radius of the beam at the intersection with the HDT beam is 48\,$\mu$m. Loading of the CDT, illustrated in stage E of Fig.~\ref{Timing}(a), involves two sub-steps. First, the power of the HDT is ramped down linearly to half of its maximum power, while simultaneously ramping up the power of the TDT beam to maximum in 150\,ms. After waiting 200\,ms in this trap, the HDT power is ramped further down in 125\,ms to 0.2\,times the maximum power, during which the power of the TDT beam is kept unchanged. This completes the loading of the CDT, at the end of which we obtain up to 1.6$\times$10$^{5}$ atoms at a temperature of 22\,$\mu$K. At this stage the peak density and PSD of the atomic ensemble are $3.7\times$10$^{12}$\,cm$^{-3}$ and 2$\times$10$^{-4}$, respectively. Next, forced evaporation in the CDT is carried out by ramping down the power of both laser beams simultaneously in seven linear steps with fixed slopes as shown in Fig.~\ref{Timing}(a) F and (b). Depending on the overall duration of the evaporation ramp $t_{\textrm{ev}}$, this phase ends with different final values of the trap depth. For an evaporation time slightly longer than 1\,s, an ensemble with 3000 condensed atoms is obtained. In our present set-up this number is mostly limited by trap laser power and insufficient vacuum conditions (background pressure $\approx 10^{-9}\,$mBar).

\begin{figure}
\centering
\includegraphics[scale=0.65, angle=0, origin=c]{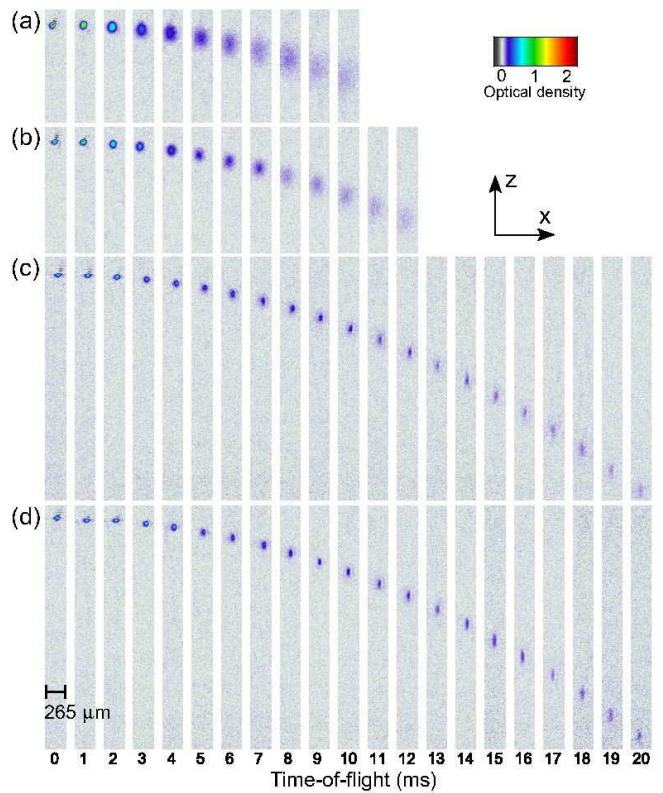}\caption{(Color online) Four series of time-of-flight images taken upon terminating the forced evaporation ramp shown in stage F of Fig.~\ref{Timing} after (a) 375\,ms, (b) 695\,ms, (c) 1015\,ms, and (d) waiting another 100\,ms in the final evaporation step. The time of free expansion is indicated below the figure. The trap depths corresponding to each series are 900\,nK, 332\,nK, 104\,nK and 104\,nK, respectively. The anisotropic expansion of the condensate fraction is clearly visible in the last two series, while in the first series an isotropically expanding thermal cloud is observed.}
\vspace{-5.mm}
\label{TOF}
\end{figure}

\begin{figure*}[!t]
\centering
\includegraphics[scale=0.3, angle=0, origin=c]{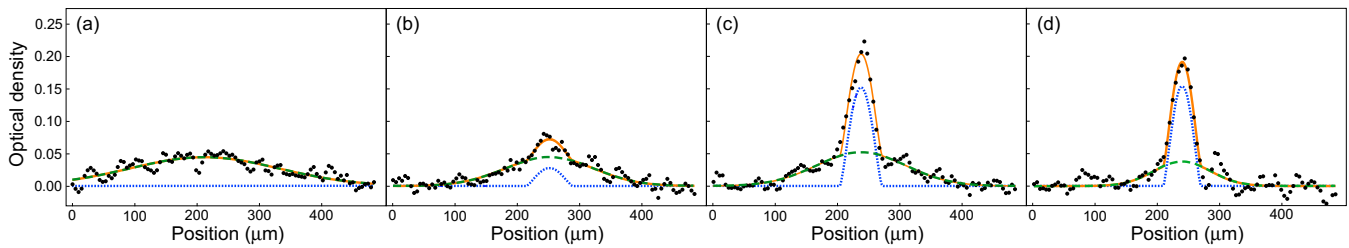}
\caption{(Color online) Column density distributions of atoms along the loose trap axis, corresponding to the $x$-axis in the series (a)-(d) of Fig.~\ref{TOF}, after 10\,ms time-of-flight. The black dots are the data points each obtained as the average over 3 measurements. The solid orange line is a bimodal fit using the sum of a Gaussian distribution (dashed green line) and a Thomas-Fermi distribution (dotted blue line).}
\vspace{-5.mm}
\label{profiles}
\end{figure*}

\begin{figure}
\centering
\includegraphics[scale=0.3, angle=0, origin=c]{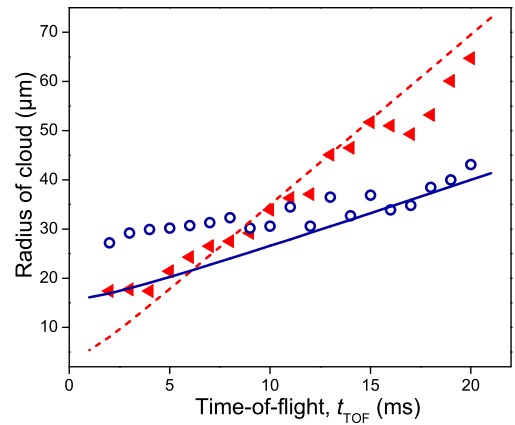}
\vspace{-6.mm}
\caption{(Color online) Thomas-Fermi radii of the condensate (corresponding to Fig.~\ref{TOF}(d)) along the strongly confined (red solid triangles) and weakly confined (blue open circles) trap axes plotted against the time-of-flight. The radii have been obtained through fits of the kind shown in Fig.~\ref{profiles}.}
\vspace{-6.mm}
\label{BEC_radii}
\end{figure}

For characterizing the cold atomic cloud, the evaporation time (and consequently the final trap depth) is varied, all CDT beams are then switched off, the atomic ensemble is allowed to expand ballistically for a variable time-of-flight (TOF) $t_{\textrm{TOF}}$ (Fig.~\ref{Timing}(a), G), and, finally, an absorption image is recorded (Fig.~\ref{Timing}(a), H). Fig.~\ref{TOF} shows four series of absorption images, in order of increasing evaporation times $t_{\textrm{ev}}$. We fit a bimodal distribution to a single column along the y-axis of each of these images, as shown in Fig.~\ref{profiles} for the images at 10\,ms time-of-flight. We obtain the temperature of the ensemble from the Gaussian part of the fits. In Fig.~\ref{TOF}(a) and Fig.~\ref{profiles}(a), corresponding to a final trap depth of 900\,nK, we observe an isotropically expanding thermal cloud with 1.9$\times$10$^{4}$ atoms at a temperature of 340\,nK. In Fig.~\ref{TOF}(b) corresponding to a trap depth of 327\,nK, a small anisotropy appears in the expansion. The bimodal fit (Fig.~\ref{profiles}(b)) clearly shows a small condensate fraction. This ensemble at a temperature of 106\,nK, contains 9200 atoms of which about 10$\%$ are condensed. We may compare this temperature to the expected critical temperature $T$$_{\mathrm{C}}$. Following \cite{Gio1996}, $T$$_{\mathrm{C}}$ for a trapped interacting Bose gas is given by:
\begin{equation}
T_{\mathrm{C}}\simeq0.94\frac{\hbar \omega N^{1/3}}{k_{B}}(1-0.73\frac{\overline{\omega}}{\omega}N^{-1/3}
-1.33\frac{a}{a_{\mathrm{HO}}}N^{1/6})
\nonumber
\end{equation}
where $\omega$=$(\omega_{1}\omega_{2}\omega_{3})^{1/3}$ and $\overline{\omega}$=$(\omega_{1}+\omega_{2}+\omega_{3})$/3 are the geometric and arithmetic means of the trap frequencies along the three axes, $N$ is the number of atoms, $a$ is the s-wave scattering length, and $a_{\textrm{HO}}=\sqrt{\hbar/m \omega}$ is the mean harmonic oscillator length. The first term gives $T$$_{\mathrm{C}}$ for a non-interacting ideal gas, whereas the second and third terms are corrections accounting for the finite particle number and interactions, respectively. By using $\omega_{1}$=2$\pi$$\times$71\,Hz, $\omega_{2}$=2$\pi$$\times$202\,Hz and $\omega_{3}$=2$\pi$$\times$174\,Hz and $a = 440\,a_{0}$ \cite{Kra2009}, we calculate $T_{\mathrm{C}}$ to be 110\,nK, which is very close to the measured value. Note that the second correction amounts to 10$\%$ of the uncorrected value, which emphasizes the dominant role of interactions for $^{40}$Ca. Figure~\ref{TOF}(c) and ~\ref{profiles}(c) shows TOF images and the column density along the y-axis at $t_{\textrm{TOF}} = 10\,$ms for a cloud cooled well below $T$$_{\mathrm{C}}$. The trap depth at this stage is 104\,nK and the harmonic frequencies are $\omega_{1}$=2$\pi$$\times$35\,Hz, $\omega_{2}$=2$\pi$$\times$181\,Hz and $\omega_{3}$=2$\pi$$\times$135\,Hz. There are 5800 atoms with 40$\%$ in the condensed fraction. After waiting for 100\,ms in this trap, so that the hottest thermal atoms are allowed to escape, we obtain the TOF-images in Fig.~\ref{TOF}(d) with the corresponding bimodal fits shown in Fig.~\ref{profiles}(d). There are now 4800 atoms, with an increased condensate fraction of about 55$\%$. In addition to the long expansion times and the clearly visible bimodal distributions in Fig.~\ref{profiles} a commonly used indication for having reached quantum degeneracy is the anisotropic expansion dynamics of the cold atomic cloud. This is further illustrated in Fig.~\ref{BEC_radii}, where the observed Thomas-Fermi radii of the condensate fraction along the tightly and weakly confined trap axes (filled triangles and open circles respectively) are plotted against $t_{\textrm{TOF}}$. The lines in the figure show the expansion expected by solving the Gross-Pitaevskii equation in the Thomas-Fermi limit following Ref.~\cite{Cas1996} with the initial radii calculated by means of the known trap frequencies, the particle number, and the scattering length. The agreement with the data is acceptable for long times-of-flight. At shorter times the data points acquire larger errors from the imperfect discrimination of the thermal cloud and the condensate fraction in the fitting procedure and limited spatial resolution of the imaging system. Furthermore, the harmonic approximation made for the trap potential underestimates the initial radii.

To conclude, we have demonstrated the second calcium BEC worldwide and provide a data basis leaving no doubt that quantum degeneracy has been reached. This has been achieved using a novel method to prepare ultracold AEM-samples, which avoids the complexities of narrow-line cooling, magic wavelength dipole traps, and laser bandwidth shaping. The method is versatile and robust, and it can be used for any state with sufficiently long lifetime (i.e., not necessarily the ground state), which can be loaded via optical pumping.

\begin{acknowledgments}
We thank P. Soltan-Panahi and J. Klinner for helpful discussions. We also thank P. S.-P. for his critical reading of the manuscript. This work was partially supported by DFG (He2334/9-1).
\end{acknowledgments}

\end{document}